\documentclass[twocolumn,showpacs,preprintnumbers,superscriptaddress,aps]{revtex4-1}

\usepackage{graphicx,color}% Include figure files

\usepackage{dcolumn}% Align table columns on decimal point

\usepackage{bm}% bold math
\usepackage{here}
\usepackage[]{lineno}

\begin{document}

\preprint{}

\title{Large anomalous Hall effect and unusual domain switching \\
in an orthorhombic antiferromagnetic material NbMnP}

\author{Hisashi Kotegawa}
\affiliation{Department of Physics, Kobe University, Kobe, Hyogo 657-8501, Japan}

\author{Yoshiki Kuwata}
\affiliation{Department of Physics, Kobe University, Kobe, Hyogo 658-8530, Japan}

\author{Vu Thi Ngoc Huyen}
\affiliation{Center for Computational Materials Science, Institute for Materials Research, Tohoku University, Sendai, Miyagi 980-8577, Japan}

\author{Yuki Arai}
\affiliation{Department of Physics, Kobe University, Kobe, Hyogo 658-8530, Japan}

\author{Hideki Tou}
\affiliation{Department of Physics, Kobe University, Kobe, Hyogo 658-8530, Japan}

\author{Masaaki Matsuda}
\affiliation{Neutron Scattering Division, Oak Ridge National Laboratory, Oak Ridge, Tennessee 37831, USA}

\author{Keiki Takeda}
\affiliation{Muroran Institute of Technology, Muroran, Hokkaido 050-8585, Japan}

\author{Hitoshi Sugawara}
\affiliation{Department of Physics, Kobe University, Kobe, Hyogo 658-8530, Japan}

\author{Michi-To Suzuki}
\affiliation{Center for Computational Materials Science, Institute for Materials Research, Tohoku University, Sendai, Miyagi 980-8577, Japan}
\affiliation{Center for Spintronics Research Network, Graduate School of Engineering Science, Osaka University, Toyonaka, Osaka 560-8531, Japan
}

\date{\today}

\begin{abstract}
Specific antiferromagnetic (AF) spin configurations generate large anomalous Hall effects (AHEs) even at zero magnetic field through nonvanishing Berry curvature in momentum space.
In addition to restrictions on AF structures, suitable control of AF domains is essential to observe this effect without cancellations among its domains; therefore, compatible materials remain limited. 
Here we show that an orthorhombic noncollinear AF material, NbMnP, acquired AF structure-based AHE and controllability of the AF domains.
Theoretical calculations indicated that a large Hall conductivity of $\sim230$ $\Omega^{-1}$cm$^{-1}$ originated from the AF structure of NbMnP.
Symmetry considerations explained the production of a small net magnetization, whose anisotropy enabled the generation and cancellation of the Hall responses using magnetic fields in different directions.
Finally, asymmetric hysteresis in NbMnP shows potential for development of controllability of responses in AF materials.
\end{abstract}

\maketitle

\section*{Introduction}

Anomalous Hall effect (AHE) is induced by an anomalous velocity of electrons perpendicular to an applied electric field through spin-orbit interactions \cite{Karplus}. 
During the last decades, the intrinsic contribution of AHE was deeply understood using the Berry-phase concept \cite{Jungwirth02,Nagaosa10}, which clarifies that the AHE is not governed by the magnetization of materials but by a geometrical effect in momentum space.
This paradigm shift facilitated symmetry analysis to yield AHE, inducing a proposal of AHE in antiferromagnetic (AF) materials \cite{Chen14,Kuber}. 
Observations of a large AHE in hexagonal systems Mn$_3$$Z$ ($Z=$Sn, Ge) at zero magnetic fields opened an avenue to develop new types of responses in AF materials \cite{Nakatsuji2015,Kiyohara16,Nayak16}, followed by observations of the anomalous Nernst effect, the magneto-optical Kerr effect, and the spin Hall effect \cite{Li17,Ikhlas17,Higo18,Kimata19}.
A key point why a large AHE occurs in these systems is that the magnetic symmetry of the AF state is the same as that of a ferromagnetic (FM) state \cite{Suzuki17}, and symmetrical conditions have been classified using the magnetic point group \cite{Smejcal20}.
This classification suggests that the AHE in AF materials occurs in various crystal and magnetic structures, regardless of whether they are noncollinear or collinear.
However, such systems are still limited, particularly for observation at zero magnetic fields, because suitable alignments of AF domains are needed to avoid cancellations among the domains. 
This alignment is a crucial issue when observing responses arising from AF structures.
A convenient way to control the AF domains is by coupling a weak spontaneous net magnetization behind the AF structure with an external magnetic field \cite{Nakatsuji2015,Kiyohara16,Nayak16,Ghimire18,Akiba20,Park22}.
Generally, such weak net magnetization is considered to be induced by the Dzyaloshinskii-Moriya (DM) interaction and geometrical frustration.
In fact, the AHE in AF materials at zero fields has been observed in several hexagonal systems with triangular lattices: Mn$_3$$Z$, CoNb$_3$S$_6$, CoTa$_3$S$_6$, and MnTe \cite{Nakatsuji2015,Kiyohara16,Nayak16,Ghimire18,Park22,Betancourt}.
They are not pure antiferromagnets because they show weak net magnetizations, but we call them AF materials in a wide sense.

%========FIGURE INSERTION=================================
\begin{figure*}[htb]
\includegraphics[width=13.5cm]{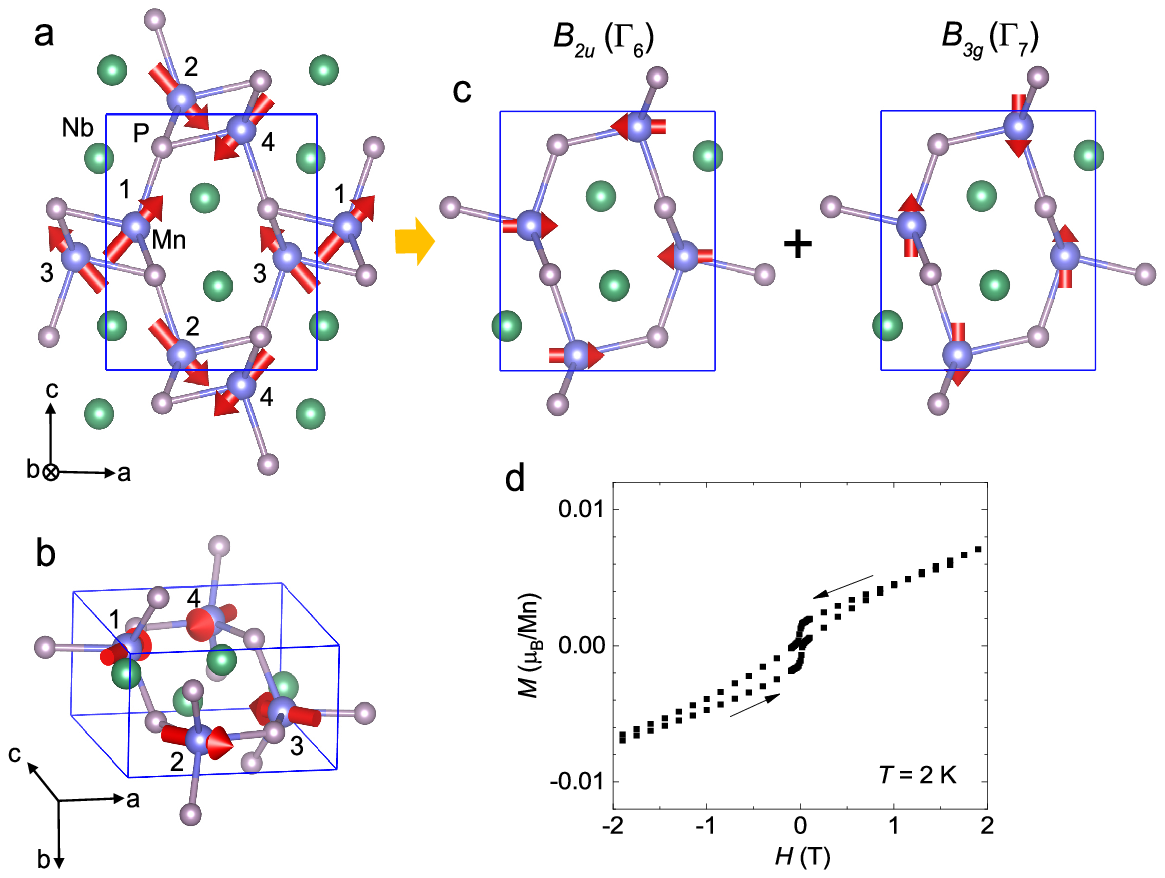}
\caption{{\bf Crystal and magnetic structures of NbMnP.} {\bf a,b} The noncollinear $Q=0$ AF structure of NbMnP \cite{Matsuda}. A unit cell (blue lines) includes four Mn atoms that are crystallography equivalent in the $Pnma$ symmetry. {\bf c} The AF structure expressed by a linear combination of odd parity $B_{2u}$ ($\Gamma_6$) and even parity $B_{3g}$ ($\Gamma_7$). $B_{3g}$ is a crucial ingredient in inducing AHE, because it possesses the same symmetry operation as the $a$-axis FM structure \cite{SM}. {\bf d} Magnetization measured at 2 K using many pieces of small single crystals \cite{Matsuda}. The observed hysteresis was broad due to the nonoriented sample.}
\label{structure}
\end{figure*}
%=========================================================

Recently, we reported that a magnetic transition with noncollinear AF structure occurs below $T_{\rm N}=233$ K in an orthorhombic system NbMnP \cite{Matsuda}.
Zhao {\it et al.} also reported similar bulk properties \cite{Zhao}.
NbMnP in $Pnma$ space group does not exhibit geometrical frustration \cite{NbMnP}, but the noncollinear AF structure is thought to be realized by competition among several exchange interactions \cite{Matsuda}.
This crystal structure possesses space-inversion symmetry, but the symmetry is broken at the center of second-, third-, and fourth-neighbor Mn--Mn bonds, which may induce DM interactions.
Nonsymmorphic NbMnP includes four equivalent Mn atoms in a unit cell.
As shown in Figs.~1a and 1b, the Mn magnetic moments of $1.2$ $\mu_B$ lie in the $ac$ plane, which were illustrated using {\it VESTA} \cite{VESTA}.
The magnetic moment of Mn1 is antiparallel to that of Mn4 and almost orthogonal to those of Mn2 and Mn3. 
This noncollinear AF structure is expressed by a linear combination of irreducible representations, odd parity $B_{2u}$ ($\Gamma_6$, the $a$-axis component) and even parity $B_{3g}$ ($\Gamma_7$, the $c$-axis component) (Fig.~1c) \cite{Matsuda}.
$B_{3g}$ corresponds to a magnetic space group $Pnm'a'$ (a magnetic point group $mm'm'$), having the same symmetry operations as the FM structure along the $a$-axis, as shown in the supplementary information \cite{SM}.
Whether FM or AF, this symmetry yields a nonzero anomalous Hall conductivity (AHC) $\sigma_{yz}$.
Here, $a$, $b$, and $c$ axes correspond to $x$, $y$, and $z$, respectively.
Behind the AF structure, a small net magnetization of a few $10^{-3}$$\mu_B$ emerges concomitantly, as shown in Fig.~1d \cite{Matsuda}, and it has been suggested to be directed along any in the $ac$ plane \cite{Zhao}.

This work reports the observed large Hall effect of an orthorhombic system, NbMnP.
The clear Hall response produced a hysteresis loop against the magnetic fields, clarifying that AHE occurred in NbMnP at zero magnetic field.
The first-principles calculation demonstrated that the AHE arose from the AF spin configuration through nonvanishing Berry curvature.
The generation of large Hall responses and their cancellations were notably controlled through domain selection by the magnetic fields in different directions.
Investigations also showed that asymmetric hysteresis, reminiscent of the exchange bias, appeared according to the direction of the magnetic field in the cooling process.
These findings in the orthorhombic structure suggest the potential controllability of noncollinear AF spin configurations and responses arising from them.

%========FIGURE INSERTION=================================
\begin{figure*}[htb]
\includegraphics[width=13.5cm]{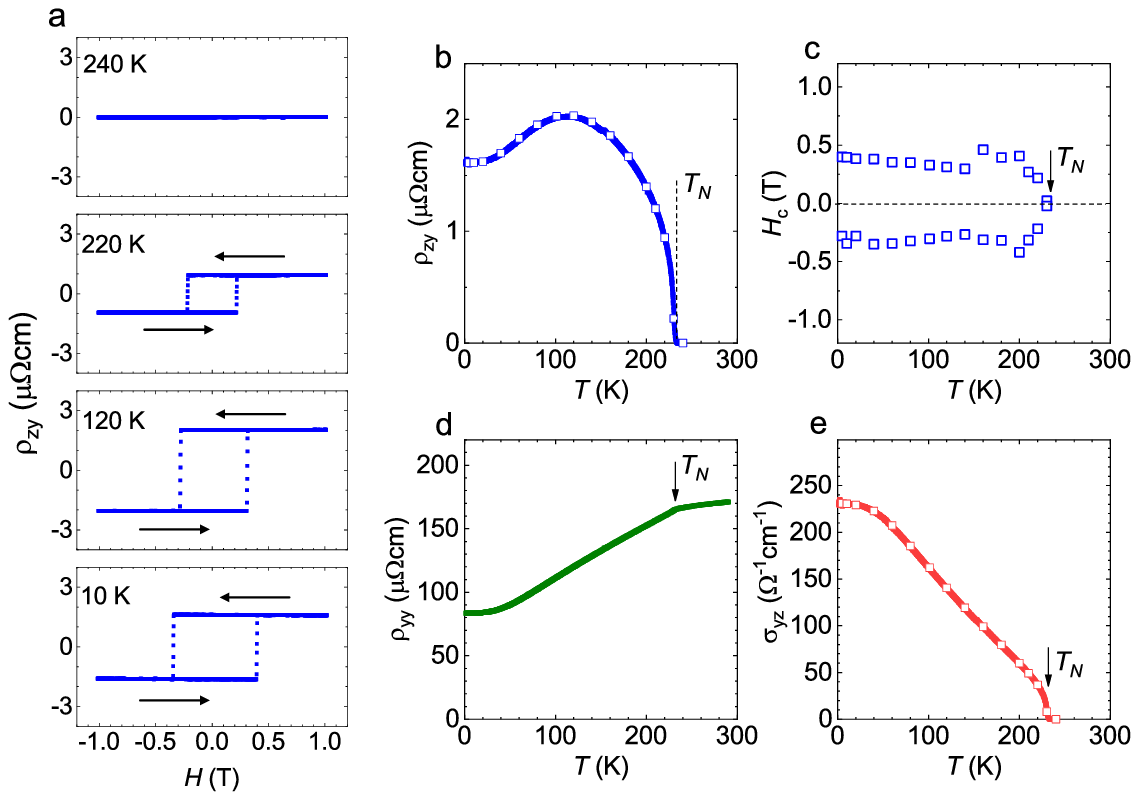}
\caption{{\bf Anomalous Hall effect in NbMnP with $T_{\rm N}=233$ K.} {\bf a} Hall resistivity, $\rho_{zy}$, against the magnetic fields along $a$-axis. The arrows indicate the field-sweep directions, and $\rho_{zy}$ shows an obvious hysteresis in an ordered state. {\bf b-e} Temperature dependences of $\rho_{zy}$, critical fields $H_c$, electrical resistivity $\rho_{yy}$, and Hall conductivity $\sigma_{yz}$ for NbMnP. Opened squares were obtained from the field sweep, while the solid line was obtained by $[ \rho_{zy}(T, H=+0.3 \ {\rm T}) - \rho_{zy}(T, H=-0.3 \ {\rm T}) ]/2$ in the temperature sweep. The maximum value $|\sigma_{yz}| \sim 230$ $\Omega^{-1}$cm$^{-1}$ was obtained below $\sim20$ K.}
\end{figure*}
%=========================================================

\section*{Results}
\subsection*{Anomalous Hall effect}

The magnetic-field dependence of the Hall resistivity $\rho_{zy}$ of NbMnP was measured under different temperatures (Fig.~2a). 
Magnetic fields were applied along the $a$-axis after a cooling at zero field. 
The data for initial application of the fields, which was done at 10 K, are not shown in the figure.
Below $T_{\rm N}=233$ K, the $\rho_{zy}$ of few $\mu \Omega$cm was observed.
Under a highly positive magnetic field, $\rho_{zy}$ was positive, and a sign change of $\rho_{zy}$ occurred, when the magnetic field increased to the negative direction.
The negative $\rho_{zy}$ returned to positive by a positive magnetic field, drawing a hysteresis loop.
At 10 K, a sign change of $\rho_{zy}$ occurred at a critical field of $H_c \simeq \pm 0.35$ T with very narrow transition width of less than a few Oe \cite{SM}.
This abrupt sign change is thought to be triggered by the inversion of the small net magnetization of a few $10^{-3}$$\mu_B$ \cite{Matsuda,Zhao}, that is, the switching of some sort of domain. 
Once the domains were aligned, the field dependence of $\rho_{zy}$ was weak, and a large value was observed even at zero field.
The ordinary Hall effect, which is proportional to the magnetic field, was negligible against the observed $\rho_{zy}$, indicating that $\rho_{zy}$ was dominated by AHE.
%The maximum value of $\rho_{yz}$ was approximately $2-3$ $\mu \Omega$cm at 120 K, which is same order as those in other AF materials showing a zero-field AHE \cite{Nakatsuji2015,Kiyohara16,Nayak16,Ghimire18,Akiba20,Park22}.
The temperature dependence of $\rho_{zy}$ is shown in Fig. 2b.
$\rho_{zy}$ rapidly increased below $T_{\rm N}$ and reached the maximum at $\sim120$ K, followed by a gradual decrease toward the lowest temperature.
Figure 2c shows the temperature dependence of $H_c$.
It was almost independent of temperature up to $\sim210$ K, above which it was reduced toward $T_{\rm N}$.
The electrical resistivity along the $b$-axis, $\rho_{yy}$, is shown in Fig.~2d, which is almost consistent with that reported by Zhao {\it et al.} \cite{Zhao}.
The $\rho_{yy}$ was insensitive to the magnetic field, as shown in the supplementary information \cite{SM}. 
The Hall conductivity $\sigma_{yz}$ was estimated through $\sigma_{yz} \simeq \rho_{zy}/(\rho_{yy}^2+\rho_{zy}^2)$ (Fig.~2e).
The $\sigma_{yz}$ continuously increased with decreasing temperature and reached the maximum of $\sim230$ $\Omega^{-1}$cm$^{-1}$ below 20 K.
This is larger than those of Mn$_3$Sn, CoNb$_3$S$_6$, and CoTa$_3$S$_6$ \cite{Nakatsuji2015,Ghimire18,Park22} and roughly half of those of Mn$_3$Ge and $\alpha$-Mn \cite{Kiyohara16,Nayak16,Akiba20}.
As the small spontaneous magnetization of a few $10^{-3}$$\mu_B$ in NbMnP generally makes it difficult to account for this large AHC \cite{Nagaosa10}, the AHE in this context was conjectured to arise from the AF structure of NbMnP through suitable AF domains.

%========FIGURE INSERTION=================================
\begin{figure*}[ht]
\includegraphics[width=13cm]{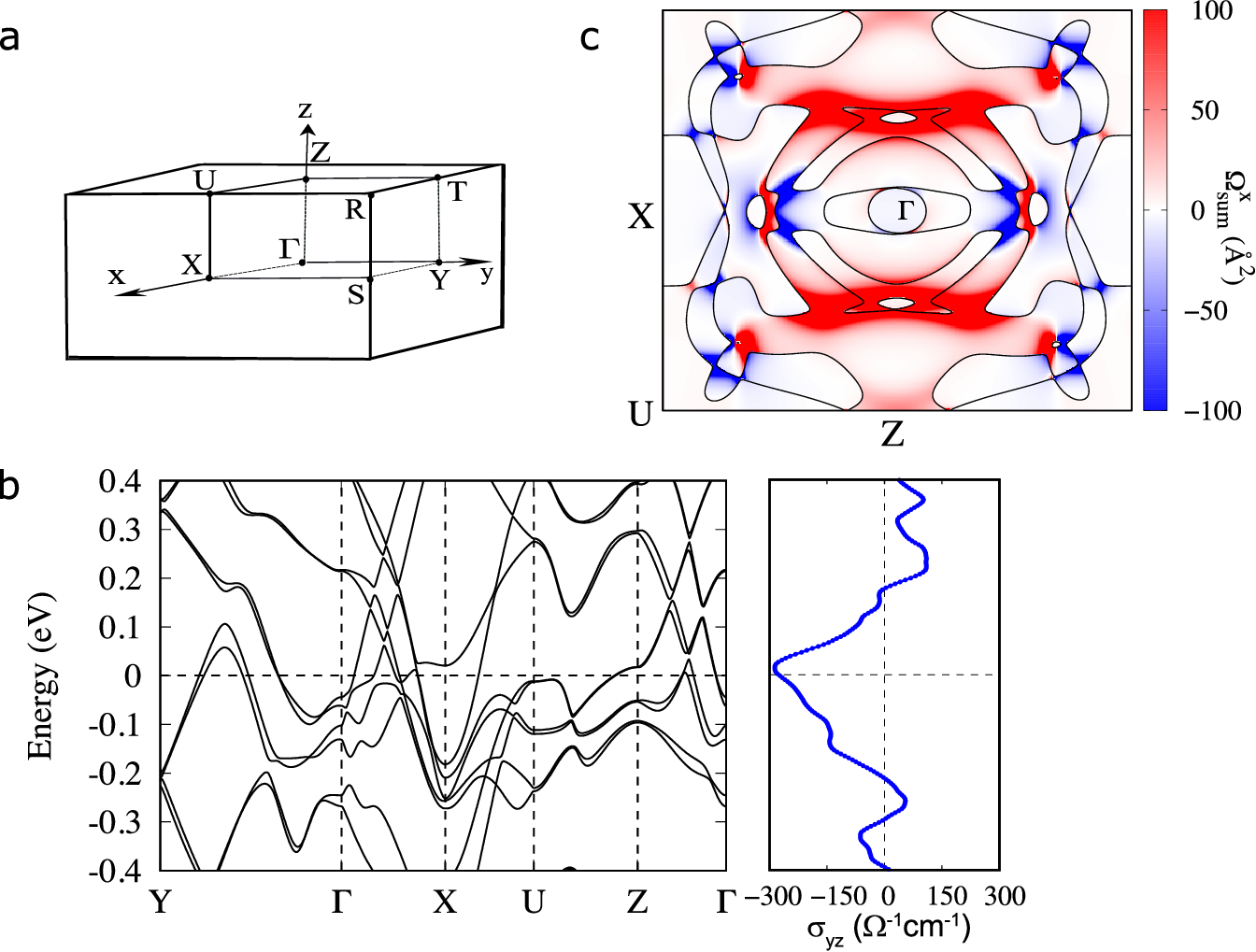}
\caption{{\bf The theoretical calculation of Berry curvature and the AHC of NbMnP.} {\bf a} BZ for the simple orthorhombic lattice. {\bf b} Calculated energy band structure for the noncollinear AF state of NbMnP. (Right panel): AHC as a function of the chemical potential. We obtained $|\sigma_{yz}| = 276$ $\Omega^{-1}$cm$^{-1}$ for the computed Fermi energy at 0 eV. The sign of $\sigma_{yz}$ depends on the AF domain.  {\bf c} The Berry curvature summed over occupied states $\Omega_{\rm sum}^x(\bm{k})$ and the Fermi surfaces (black curves) in the $k_y=0$ plane. The positive (or negative) values are displayed in red (blue). }
\label{structure}
\end{figure*}
%=========================================================

\subsection*{Calculation of anomalous Hall conductivity}

To investigate origin of the AHE, we evaluated the Berry curvature and AHC in NbMnP for the noncollinear AF spin configuration {\it via} first-principles calculations. 
Here, we set the magnetic moments at 1.2 $\mu_B$/Mn, and the direction of the magnetic moments was the opposite of Fig.~1a (the next section shows domain I\hspace{-1.2pt}I\hspace{-1.2pt}I).
Details of the calculation are described in the Methods section.
Figure 3a shows the Brillouin Zone (BZ) of NbMnP, and Fig. 3b shows the energy band structure along the high-symmetry line.
The Berry curvature summed over the occupied states, $\Omega_{\rm sum}^x(\bm{k}) = \sum_n f [ \varepsilon_n(\bm{k})-\mu ] \Omega_n^x(\bm{k})$ in the $k_y=0$ plane is shown in Fig. 3c. Here, $\Omega_n^x(\bm{k})\equiv\Omega_{n,yz}(\bm{k})$.
The $\Omega_{\rm sum}^x(\bm{k})$ exhibited a largely positive value in the intermediate region between the $\Gamma$ and $Z$ points.
The avoided crossing bands near the Fermi energy contribute to a non-zero Berry curvature as suggested for Mn$_3$$A$N \cite{Huyen}. The left panel of Fig. 3b shows the dense avoided crossing bands between $\Gamma$ and $X$ and between $\Gamma$ and $Z$. As a consequence, the Berry curvature shows strong intensity around the Fermi surfaces on these lines as seen in Fig. 3c. Figure 3c also shows strong intensity region of the positive Berry curvature is apparently  larger than the negative one and result in the large negative AHC with the Fermi energy obtained in the calculation from Eq. (1), as shown in the left panel of Fig. 3b. 
$\sigma_{yz}$ coincidentally exhibited the maximum near the Fermi level, with $|\sigma_{yz}| = 276$ $\Omega^{-1}$cm$^{-1}$ being suggested.
Our investigations revealed that this value agreed well with the experimental $|\sigma_{yz}|=230$ $\Omega^{-1}$cm$^{-1}$, confirming that the observed AHE arose from the AF structure of NbMnP.

%========FIGURE INSERTION=================================
\begin{figure*}[ht]
\includegraphics[width=15cm]{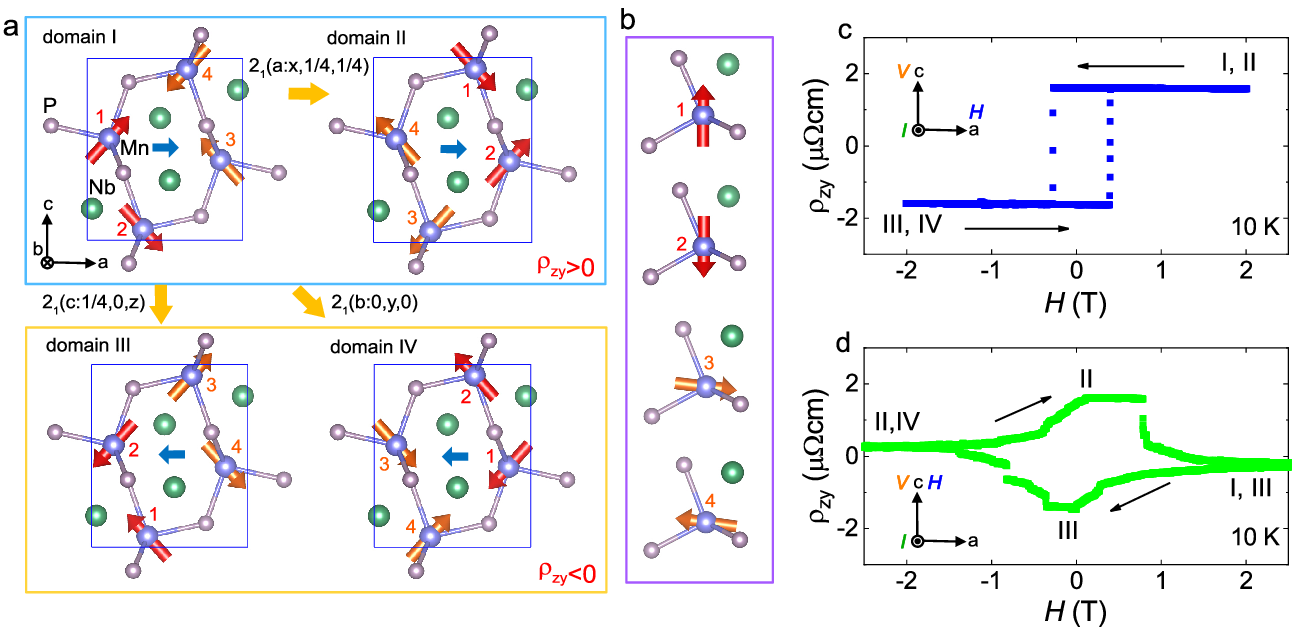}
\caption{{\bf Considerations of AF domains for NbMnP and their switching by magnetic fields.} {\bf a} The expected AF domains for NbMnP are illustrated using {\it VESTA} \cite{VESTA}. A symmetry operation $2_1$ connects the Mn sites. For example, $2_1(a: x, 1/4, 1/4)$ indicates the screw operation along the $a$-axis with the screw axis represented by $(x, 1/4, 1/4)$. The Mn atoms with same numbers connected by the symmetry operations have same sizes of the magnetic moments. The combination of $B_{2u}$ and $B_{3g}$ induces symmetry lowering, which makes Mn1 and Mn2 inequivalent to Mn3 and Mn4. For example, the magnetic moments in domain I are expressed by Mn1:$(u,0,v)$, Mn2:$(u,0,-v)$, Mn3:$(-u',0, v')$, and Mn4:$(-u',0,-v')$, where $u$, $v$, $u'$, and $v'$ are components of the magnetic moments. When we assume $u>u'$, nonzero net magnetizations along the blue arrow remain in each domain. {\bf b} Directions of the respective Mn moments against the surrounding ionic arrangements. {\bf c,d} The Hall resistivities at different magnetic-field directions. For $H \parallel a$, switching between (I and I\hspace{-1.2pt}I) and (I\hspace{-1.2pt}I\hspace{-1.2pt}I and I\hspace{-1.2pt}V) was suggested with theoretical calculations revealing domain I and I\hspace{-1.2pt}I under positive magnetic fields. For $H \parallel c$, switching between (I and I\hspace{-1.2pt}I\hspace{-1.2pt}I) and (I\hspace{-1.2pt}I and I\hspace{-1.2pt}V) was expected. In the figure, we assumed that domains I and I\hspace{-1.2pt}I\hspace{-1.2pt}I were realized under positive magnetic fields. Domain selection was induced by decreasing fields. This is considered to originate from rotation of the net magnetization and misalignment of the magnetic field. Under a positive (negative) $c$-axis magnetic field, a small negative (positive) $a$-axis magnetic field is expected. 
}
\label{structure}
\end{figure*}
%=========================================================

\subsection*{Switching of AF domains}

Consistent with the above calculation, this study observed a large AHE, indicating the selection of suitable AF domains under magnetic fields along the $a$-axis. 
As nonsymmorphic NbMnP includes four Mn atoms per unit cell, we expect four AF domains in Fig.~4a from the symmetry operations connecting the respective Mn sites.
Domains I and I\hspace{-1.2pt}I have the same $c$-axis components of Mn moments, which yield $\rho_{zy}$ with the same signs.
From the first-principles calculations, while these domains generated a positive $\rho_{zy}$,
domains I\hspace{-1.2pt}I\hspace{-1.2pt}I and I\hspace{-1.2pt}V with opposite $c$-axis components generated a negative $\rho_{zy}$.
These findings indicate weak spontaneous net magnetizations directed along the $a$-axis and opposite between the domains (I and I\hspace{-1.2pt}I) and (I\hspace{-1.2pt}I\hspace{-1.2pt}I and I\hspace{-1.2pt}V).
Considering the magnetic symmetry of NbMnP, we can explain the origin of weak net magnetizations in two ways.
The first explanation is the DM interaction, which breaks the AF coupling of the third-neighbor Mn1--Mn4 and Mn2--Mn3 bonds.
The DM vectors for these bonds are directed along the $b$-axis, because these bonds lie in the mirror plane \cite{Moriya}, canting the AF couplings in the $ac$ plane.
For the $B_{3g}$ components, this canting yields a net magnetization along the $a$-axis, because $B_{3g}$ allows FM components along the $a$-axis \cite{SM}.
The second explanation is the symmetry lowering due to a combination of two irreducible representations, $B_{2u}$ and $B_{3g}$.
Although each magnetic symmetry of $B_{2u}$ and $B_{3g}$ is allowed in crystal symmetry of $Pnma$, their combination lowers the crystal symmetry of $Pnma$ in principle, causing the common magnetic symmetry operations between $B_{2u}$ and $B_{3g}$ to remain.
This consideration indicates the crystal symmetry below $T_{\rm N}$ as the space group $Pmn2_1$ \cite{SM}, which is noncentrosymmetric and polar.
Note that a magnetic space group is $Pm'n2_1'$ and a magnetic point group is $m'm2'$.
Symmetry lowering of the crystal structure has not been observed within the experimental resolution \cite{SM}.
This symmetry lowering to $Pmn2_1$ divides the equivalent Mn sites in $Pnma$ into two different sites.
Actually, directions of the magnetic moments against the surrounding ionic arrangements are not equivalent for four Mn sites (Fig.~4b).
Those of Mn1 and Mn2 are opposite and regarded as equivalents, but they differ from Mn3 and Mn4.
Therefore, for example, the magnetic moments in the domain I are expressed by Mn1:$(u,0,v)$, Mn2:$(u,0,-v)$, Mn3:$(-u',0, v')$, and Mn4:$(-u',0,-v')$, where $u$, $v$, $u'$, and $v'$ are components of the magnetic moments.
The net magnetization along the $a$-axis is induced by $u\neq u'$, that is, breaking of the AF couplings of Mn1--Mn4 and Mn2--Mn3 bonds.
When $u\neq u'$ while keeping $u^2+v^2=u'^2+v'^2$, rotations of the magnetic moments induce the net magnetization.
This corresponds to the DM interaction mentioned above. 
Another case is $u\neq u'$ with $u^2+v^2 \neq u'^2+v'^2$.
In this case, a main cause to induce the net magnetization is a difference in the sizes of the magnetic moments, and we can interpret that it is induced by local magnetic anisotropy in the $ac$ plane.
In either case, an important ingredient is $B_{3g}$, where FM components are symmetrically allowed \cite{SM}.

In the four AF domains, the Mn sites connected by a symmetry operation are indicated by the same numbers (Fig.~4a).
The Mn sites indicated by the same numbers have the equivalent magnetic moments at zero fields.
%The breaking of equivalency between Mn1, 2 and Mn3, 4 induces nonzero net magnetizations along the $a$-axis.
Notably, symmetry considerations revealed that the net magnetizations are same directions between domains I and I\hspace{-1.2pt}I and opposite directions in domains I\hspace{-1.2pt}I\hspace{-1.2pt}I and I\hspace{-1.2pt}V \cite{SM}.
This feature enables domain switching between (I and I\hspace{-1.2pt}I) and (I\hspace{-1.2pt}I\hspace{-1.2pt}I and IV), using the $a$-axis magnetic field.
This switching also induces the inversion of the $c$-axis components of the magnetic moments related to the sign of $\rho_{zy}$.

We compare the $\rho_{zy}$ between $H \parallel a$ and $H \parallel c$ in Figs.~4c and 4d.
The $H \parallel a$ has the same configuration as those in Fig.~2 possessing the large Hall responses.
First-principles calculation suggested that while a positive $\rho_{zy}$ arose from the domains I and I\hspace{-1.2pt}I, a negative $\rho_{zy}$ came from the domains I\hspace{-1.2pt}I\hspace{-1.2pt}I and I\hspace{-1.2pt}V.
In contrast, the $\rho_{zy}$ for $H \parallel c$ was very small for high magnetic fields, indicating that the selected domains differed from $H \parallel a$.
The magnetic field along the $c$-axis breaks the symmetry of $Pmn2_1$, making all four Mn sites inequivalent.
Under this condition, the averaged $c$-axis magnetizations between the domains (I and I\hspace{-1.2pt}I\hspace{-1.2pt}I) and (I\hspace{-1.2pt}I and I\hspace{-1.2pt}V) differ \cite{SM}, indicating that positive and negative $c$-axis magnetic fields switched between the domains (I and I\hspace{-1.2pt}I\hspace{-1.2pt}I) and (I\hspace{-1.2pt}I and I\hspace{-1.2pt}V).
Both domain sets cancel $\rho_{zy}$, consistent with the small $\rho_{zy}$ at higher magnetic fields along the $c$ axis.
Experimentally, this cancellation collapsed below $\sim \pm1.5$ T, and $\rho_{zy}$ showed large values similar to $H \parallel a$, when the magnetic field went beyond zero, suggesting that decreasing fields induces the domain selection.
This behavior is explained by assuming that the net magnetization rotates toward the $c$-axis under strong $c$-axis magnetic fields.
The $a$-axis component of the net magnetization is conjectured to be only effective at low $c$-axis magnetic fields, where domain selection is possible by misalignment of the magnetic field.

%========FIGURE INSERTION=================================
\begin{figure}[htb]
\includegraphics[width=7.5cm]{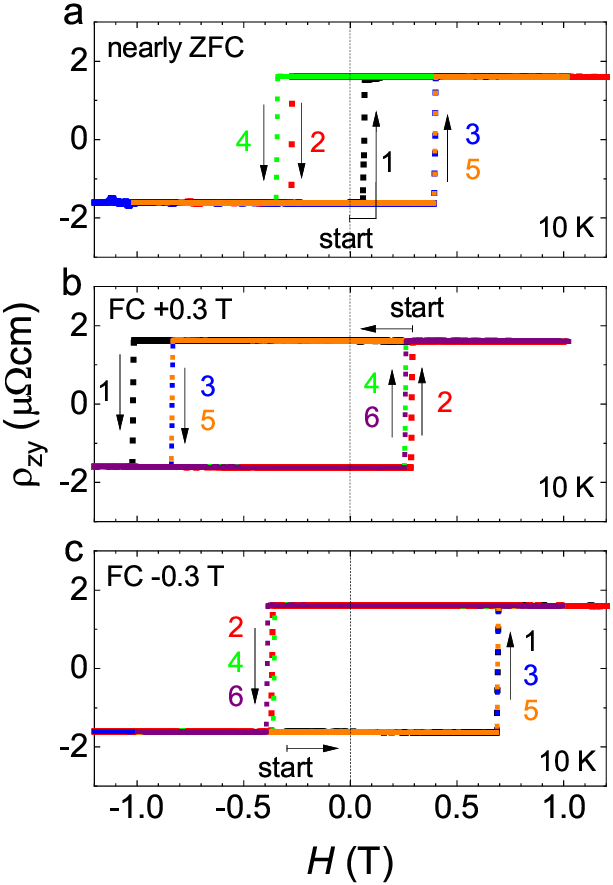}
\caption{{\bf Asymmetric hysteresis appearing after FC in NbMnP.} Hysteresis loops of $\rho_{zy}$ at 10 K after different cooling processes; {\bf a} ZFC, {\bf b} FC under $+0.3$ T, and {\bf c} FC under $-0.3$ T. The numbers indicate the order of field sweeps. The arrows of ``start'' show the sweeping direction from the initial position. Asymmetric hysteresis appeared after FC, reminiscent of the exchange bias. }
\end{figure}
%=========================================================

\subsection*{Asymmetric hysteresis}

%The $H_c$ in NbMnP was much larger than $\sim50$ mT in Mn$_3$$Z$ ($Z=$Sn, Ge) and $\alpha-$Mn \cite{Nakatsuji2015,Kiyohara16,Nayak16,Akiba20}.
%The large $H_c$ was observed in chiral AF systems CoNb$_3$S$_6$ and CoTa$_3$S$_6$ \cite{Ghimire18,Park22}.
%A notable feature in NbMnP is a difference in $H_c$ between positive and negative fields, which was remarkable in crystal B1.
%To check this feature in more detail, differences in $H_c$ were investigated after different cooling processes using the more homogeneous crystal A.
Another unusual domain switching in NbMnP was found when the cooling process was changed. 
Figure 5 shows the hysteresis loops at 10 K measured after the different processes.
Hysteresis loops were drawn several times in each setting, and they showed a good reproducibility, although $H_c$ changed occasionally within 20 \% for the magnetic fields along the same direction.
After a zero-field cooling (ZFC), the initial state showed a large negative $\rho_{zy}$, probably owing to the residual magnetic field from a superconducting magnet.
In this case, the hysteresis loop was almost symmetric against the positive and negative magnetic fields, except for the first step.
The hysteresis loops after cooling under magnetic fields of $\pm 0.3$ T (FC) are shown in Figs. 5(b) and 5(c).
The direction of the applied field on the cooling process selected the sign of the initial $\rho_{zy}$.
As the $H_c$ to inverse the initial domain was obviously enhanced, the hysteresis loop showed asymmetry.
Asymmetric hysteresis has been observed in artificial bilayers of FM and AF materials through interactions across the interface, well known as the exchange bias or the exchange anisotropy \cite{Meiklejohn,Nogues,Berkowitz}.
This phenomena has been used in magnetic storage.
Although a key point of this effect is the robustness of the magnetic moments in the AF layer against magnetic fields, this is not the case for bulk crystal NbMnP.
On the analogy of the exchange bias, it was conjectured that magnetic moments in some parts of NbMnP were pinned regardless of the flipped domains.
Therefore, leaving aside the detailed mechanism, spatially separated electronic states could be formed at regions near the deficiencies of Nb atoms with a few percentage \cite{SM} or at the domain wall. 
As for the domain wall, even though $\rho_{zy}$ was highly positive or negative, the crystal was always expected to include two domains, either a pair of (I and I\hspace{-1.2pt}I) or (I\hspace{-1.2pt}I\hspace{-1.2pt}I and I\hspace{-1.2pt}V); therefore, the domain wall was inevitably present. 
When the domain switching occurs between a pair connected by time-reversal symmetry (I $\leftrightarrow$ I\hspace{-1.2pt}I\hspace{-1.2pt}I and I\hspace{-1.2pt}I $\leftrightarrow$ I\hspace{-1.2pt}V) without exchanging the inequivalent Mn sites, the domain wall does not need to move regardless of the flipped domains.
However, it is an open question whether such regions, which are probably small, can store pinned moments to stabilize an initial domain.
Accordingly, microscopic investigations, such as direct observation of the AF domains, should solve this mechanism.
In any case, a difference from the conventional exchange bias is that the flipped domain in NbMnP is the AF domain.
This unexpected feature shows that asymmetric hysteresis occurred in bulk AF materials without any artificial interface, which is promising for the future development of controllability of AF domains.

\section*{Discussion}

In this study, noncollinear AF system, NbMnP, with an orthorhombic structure showed a zero-field AHE, reinforcing the validity of the symmetry analysis for occurrences of AHE.
The large AHC of $\sigma_{yz}\sim230$ $\Omega^{-1}$cm$^{-1}$ and quantitative consistency with the theoretical calculation demonstrated that it arose from the AF structure of NbMnP.
The symmetry considerations explained that a weak net magnetization was produced by the DM interaction or the symmetry lowering due to two irreducible representations, being the key ingredient in aligning the AF domains. 
Furthermore, investigations showed that due to the magnetic anisotropy in the $ac$ plane of the orthorhombic structure, domain selections depended on magnetic-field directions; the $a$-axis magnetic field produced large Hall responses that changed its sign, and the $c$-axis magnetic field induced the cancellation of the Hall responses. 
Although contribution of orthorhombicity is unclear, the asymmetric hysteresis observed in bulk crystals shows potential for the development of the controllability of AF domains. 
These findings demonstrate that various crystal systems can induce large Hall responses even at zero magnetic field, generating new insights for studies of AHE and future applications of AF materials.

A notable perspective other than the AHE in NbMnP is that the odd parity $B_{2u}$ expression includes the magnetic toroidal dipole moment.
The $Q=0$ magnetic ordering in NbMnP generates a ferroic magnetic toroidal dipole ordering, which can induce cross-correlated couplings such as a magnetoelectric effect \cite{Hayami14,Saito18,Ohta22}, and current-induced domain switching \cite{Watanabe}.
The domain selectivity using the $a$- and $c$-axes magnetic fields in NbMnP is expected to produce a single-domain ferroic ordering state, which can assist to generate large responses.

\section*{Methods}
\subsection*{Single crystal growth}
Single crystals of NbMnP were grown using the self-flux method \cite{Matsuda}.
The starting materials, comprising Nb powder, Mn powder, and P flakes (molar ratio: 1.25 : 85.9 : 12.85), were first placed in an Al$_2$O$_3$ crucible, and the set-up was sealed in an evacuated quartz ampoule. 
Next, we gradually heated the ampoule to 1200 ${\rm ^o C}$ and held it at this temperature for 6 hours, followed by slow cooling to 900 ${\rm ^o C}$ at $-3.3$ ${\rm ^o C}$/h.
After the excess substances had been decomposed using a diluted nitric acid solution, single crystals of needle-like shape were obtained.
Crystal symmetry, lattice parameters, and the occupancy of each site were checked by single-crystal X-ray diffraction measurements using a Rigaku Saturn724 diffractometer.
The Laue diffraction method determined the crystal axis' direction.
Consequently, our single crystals were needle-like along the $b$-axis, whereas the typical width along the $a$- and $c$-axes was $\sim0.2$ mm.

\subsection*{Hall resistivity measurements}

First, we shaped a crystal for the Hall measurements by polishing it with sandpaper. 
Using the spot-weld method, we made electrical contacts for gold wires (25 $\mu$m in diameter). 
Then, we measured the Hall resistivity using a standard four-probe method and an AC resistance bridge (Model 370, Lake Shore). 
Finally, we antisymmetrized the Hall resistivity against magnetic fields to remove the longitudinal component induced by contact misalignment. 
We estimated the angle accuracy of the crystal axis to be $<5$$^\circ$ with respect to applying magnetic fields.

\subsection*{First-principles analysis of AHC}

First-principles calculations were performed using the QUANTUM ESPRESSO package \cite{Giannozzi}.
The generalized gradient approximation (GGA) in the parametrization of Perdew, Burke, and Ernzerhof \cite{Perdew} was used for the exchange-correlation functional, and the pseudopotentials in the projector augmented-wave method \cite{Blochl,Kresse} were generated by PSLIBRARY \cite{Corso}. 
The lattice constants $a=6.1661 {\rm \AA}, b=3.5325 {\rm \AA}$, and $c=7.2199{\rm \AA}$ from previous experiments at 9 K~\cite{Matsuda}
were used. 
Starting from the experiment atomic positions, until residual forces $<0.01$ eV/{\rm\AA} were reached, the atomic positions were fully relaxed. 
Next, we chose kinetic cutoff energies of 50 and 400 Ry as the plane-wave basis set and charge density, respectively. 
A $k$ mesh of $9 \times 15 \times 9$ had sampled the first BZ with a Methfessel-Paxton smearing width of 0.005 Ry to get the Fermi level. 
We used the PAOFLOW package~\cite{Naredelli,Cerasoli} to estimate AHC. 
A 18 $\times$ 30 $\times$ 18 Monkhorst-Pack $k$-point grid generated a tight-binding set of pseudo-atomic orbitals for subsequent AHC calculations.
Wannier90 \cite{Giovanni} plotted the Berry curvature in the BZ. 
The $4d$ and $5s$ orbitals of Nb, the $3d$ and $4s$ orbitals of Mn, and the $3s$ and $3p$ of P were included for the Wannier interpolation scheme using Wannier90 to construct realistic tight-binding models from the first-principles band structures \cite{Wang}.

The AHC was calculated using the Kubo formula \cite{Wang}.
\begin{equation}
\sigma_{\alpha \beta} = -\frac{e^2}{\hbar} \int \frac{d \bm{k}}{(2 \pi)^3} \sum_n f [ \varepsilon_n(\bm{k})-\mu ] \Omega_{n,\alpha \beta}(\bm{k}),
\end{equation}
where $n$ is the band index and $\alpha, \beta = x, y,$ and $z$ with $\alpha \neq \beta$.
The Berry curvature is defined as
\begin{equation}
\Omega_{n,\alpha \beta}(\bm{k}) = - 2 {\rm Im} \sum_{m\neq n} \frac{v_{nm,\alpha}(\bm{k}) v_{mn,\beta}(\bm{k}) }{[ \varepsilon_m(\bm{k})-\varepsilon_n(\bm{k}) ]^2}.
\end{equation}
The $\varepsilon_n(\bm{k})$ is the eigenvalue, and 
\begin{equation}
v_{nm,\alpha}(\bm{k}) = \frac{1}{\hbar} \left< u_n(\bm{k}) \left| \frac{\partial \hat{H}(\bm{k})}{\partial k_{\alpha}} \right| u_m(\bm{k}) \right>,
\end{equation}
where $u_n(\bm{k})$ is the periodic-cell part of the Bloch states and $\hat{H}(\bm{k}) = e^{-ik \cdot r} \hat{H} e^{ik \cdot r}$.

\section*{Acknowledgments}
We thank Hisatomo Harima and Youichi Yanase for valuable discussions and comments. This work was supported by JSPS
KAKENHI Grant Nos. 18H04320, 18H04321, 19H01842, 21H01789, 21H04437, 21K03446, and 23H04871, Iketani Science and Technology Foundation, Hyogo Science and Technology Association, and The Murata Science Foundation.
A part of the numerical calculation was performed in MASAMUNE-IMR of the Center for Computational Materials Science, Institute for Materials Research, Tohoku University.

\section*{Author contributions}
H.K. conceived of and designed the study. 
H.K., Y.K, and H.S. synthesized the single crystal.
K.T. performed the single-crystal X-ray diffraction measurements. 
H.K., Y.A., and H.T. performed the Hall resistivity measurements.
V.T.N.H. and M.T.S. calculated the Berry curvature and AHC, and provided theoretical input for the interpretation of the results.
M.M. provided information about the magnetic symmetry obtained from the neutron scattering experiments.
H.K. wrote the paper with assistance from V.T.N.H. and M.T.S.

\section*{Competing interests}
The authors declare no competing interests.

\section*{Data Availability}
The data used in this study are available from the corresponding author upon request.

\clearpage

\setcounter{figure}{0}

\noindent
{\bf Supplementary information: Large anomalous Hall effect at zero magnetic field and unusual domain switching in an orthorhombic antiferromagnetic material NbMnP
}

\subsection{Magnetic structure symmetry}

The $Q=0$ noncollinear magnetic structure in NbMnP is represented by a linear combination of $B_{2u}$ ($\Gamma_6$) and $B_{3g}$ ($\Gamma_7$).
Table I shows allowed magnetic moments at four Mn sites for each symmetry.
The neutron scattering data were reproduced by the sum of the $a$-axis components of $B_{2u}$ and the $c$-axis components of $B_{3g}$ \cite{Matsuda1}.
Here, the other components were neglected, and its antiferromagnetic (AF) structure is represented by $B_{2u}+B_{3g}$ (AF). 
If the symmetry-allowed components are nonzero, the magnetic structure is represented by $B_{2u}+B_{3g}$ (SA), where the Mn sites are divided into two.

The $a$-axis components for $B_{3g}$ correspond to an FM structure.
Even though the $a$-axis component is zero, any magnetic structure in $B_{3g}$ is in the same symmetry as the FM structure along the $a$-axis; therefore, a nonzero Hall conductivity $\sigma_{yz}$ emerges.
The collinear $B_{3g}$ component of NbMnP is classified as a magnetic toroidal quadrupole.

\begin{table*}[htb]
\caption{The $Q=0$ magnetic structure for the irreducible representation (IR) of NbMnP. $u$ and $v$ represent the components of the magnetic moments. The positions of four Mn atoms are defined as follows: Mn1: ($x, y, z$), Mn2: ($-x$+1/2, $-y$, $z$+1/2), Mn3: ($-x$, $y$+1/2, $-z$), and Mn4: ($x$+1/2, $-y$+1/2, $-z$+1/2).
$\dot{u}$ and $\dot{v}$ are negligible components for the analysis in the neutron scattering data, and the AF structure $B_{2u}+B_{3g}$ (AF) is obtained through $\dot{u}=0$ and $\dot{v}=0$ \cite{Matsuda1}. When the symmetry-allowed components $\dot{u}$ and $\dot{v}$ are nonzero, the magnetic moments in $B_{2u}+B_{3g}$ (SA: symmetry allowed) are obtained through conversions of $u+\dot{u} \rightarrow u$, $v+\dot{v} \rightarrow v$, $u-\dot{u} \rightarrow u'$, and $v-\dot{v} \rightarrow v'$.
}
\begin{center}
\begin{tabular}{cccccc}\hline
IR & &Mn1 & Mn2 & Mn3 & Mn4 \\ \hline
$B_{2u}$ ($\Gamma_6$) & & ($u$, 0, $\dot{v}$) & ($u$, 0, $-\dot{v}$) & ($-u$, 0, $-\dot{v}$) & ($-u$, 0, $\dot{v}$)\\
$B_{3g}$ ($\Gamma_7$) & & ($\dot{u}$, 0, $v$) & ($\dot{u}$, 0, $-v$) & ($\dot{u}$, 0, $v$) & ($\dot{u}$, 0, $-v$)\\ \hline
& & & & & \\ \hline
$B_{2u}+B_{3g}$ (AF) & & ($u$, 0, $v$) & ($u$, 0, $-v$) & ($-u$, 0, $v$) & ($-u$, 0, $-v$)\\
$B_{2u}+B_{3g}$ (SA) & & ($u$, 0, $v$) & ($u$, 0, $-v$) & ($-u'$, 0, $v'$) & ($-u'$, 0, $-v'$)\\
\hline
\end{tabular}
\end{center}
\end{table*}

Table I\hspace{-1.2pt}I shows symmetry operations for each irrespective representation.
The $Pnma$ space group possesses eight symmetry operations. 
The $E$ means identity operation, and $2_1$ indicates screw operation along each axis.
The $i$ is the space-inversion symmetry. 
The $n$, $m$, and $a$ correspond to $n$-glide, mirror, and $a$-glide operations, respectively. 
The number $-1$ corresponds to time-reversal symmetry required to reproduce directions of the magnetic moments. 
Different signs between $B_{2u}$ and $B_{3g}$ in a specific symmetry operation indicate competition of symmetry operations, that is, the symmetry is broken in the magnetically ordered state of NbMnP.
Therefore, remained symmetry operations below $T_{\rm N}$ are $E$, $2_1[001]$, $n$, and $m$ among the eight operations.
This corresponds to the space group $Pmn2_1$ (or $Pnm2_1$) among maximum subgroups of $Pnma$ \cite{IT}.
The space group $Pmn2_1$ is noncentrosymmetric and polar.
This symmetry breaking divides the Mn sites into two kinds of Mn sites.
This corresponds to $B_{2u}+B_{3g}$ (SA) in Table. I, where the magnetic moments differ between two kinds of Mn sites.

The magnetic structure expressed by the $B_{3g}$ component corresponds to the magnetic space group $Pnm'a'$.
This belongs to the magnetic point group of $mm'm'$.
This satisfies the condition of a nonzero Hall conductivity \cite{Smejcal20a}.
If we recognize the symmetry lowering to $Pmn2_1$ due to the mixing of the $B_{3g}$ and $B_{2u}$ components, the magnetic space group is $Pm'n2_1'$.
This is in the magnetic point group $m'm2'$, which also induces a nonzero Hall conductivity \cite{Smejcal20a}.

%========TABLE INSERTION=================================
\begin{table}[htb]
\caption{Symmetry operations for two IRs in the $Pnma$ space group. The number $-1$ indicates the time-reversal symmetry required to reproduce directions of the magnetic moments. }
\label{t2}
\raggedright
\begin{center}
\begin{tabular}{ccccccccc}
\hline
IR&$E$ & $2_1[100]$ & $2_1[010]$ & $2_1[001]$ & $i$ & $n[100]$ & $m[010]$ & $a[001]$ \\
\hline
$B_{2u}$ ($\Gamma_6$) & 1 & $-1$ & 1 & $-1$ & $-1$ & 1 & $-1$ & 1 \\
$B_{3g}$ ($\Gamma_7$) & 1 & 1 & $-1$ & $-1$ & 1 & 1 & $-1$ & $-1$ \\
\hline
\end{tabular}\\
\end{center}
\end{table}
%=========================================================

\newpage

\subsection{Transition width of the domain switching}

The field-swept data in the main text were obtained by sweeping magnetic fields with a rate of 0.088 T/min.
Intermediate points between positive $\rho_{yz}$ and negative $\rho_{yz}$ were plotted in the figure, but most points are unlikely intrinsic, because they were recorded owing to an average function of a resistance bridge.
Figures 1(a) and 1(b) show transitions for careful measurements with a slow sweeping rate of 0.0088 T/min.
The transition width was less than a few Oe, and intermediate points were not recorded.
The inversion of magnetic domains occurred simultaneously in almost all parts of the measured crystal.

%========FIGURE INSERTION=================================
\begin{figure}[htb]
\includegraphics[width=8cm]{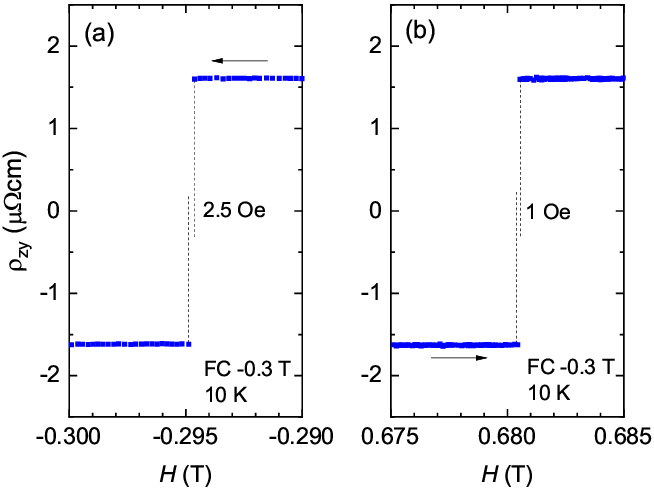}
\caption{(a), (b) The sign change of $\rho_{yz}$ observed with a slow sweeping rate. The transition occurred abruptly in almost all parts of the measured crystal.}
\label{structure}
\end{figure}
%=========================================================

\newpage

\subsection{Field dependence of the electrical resistivity}

To confirm that the observed Hall resistivity arose from a component normal to the current, we measured the field dependence of $\rho_{yy}$.
The raw data for $\rho_{yy}$ are shown in Fig.~2.
The very small hysteresis appeared, but they were $\sim4$\% of anomalies in $\rho_{yz}$.
It is reasonable to consider that this anomaly was induced by small misalignment of the electrical contacts.
The field-dependences of intrinsic $\rho_{yy}$ are very small and negligible.

%========FIGURE INSERTION=================================
\begin{figure}[htb]
\includegraphics[width=6.5cm]{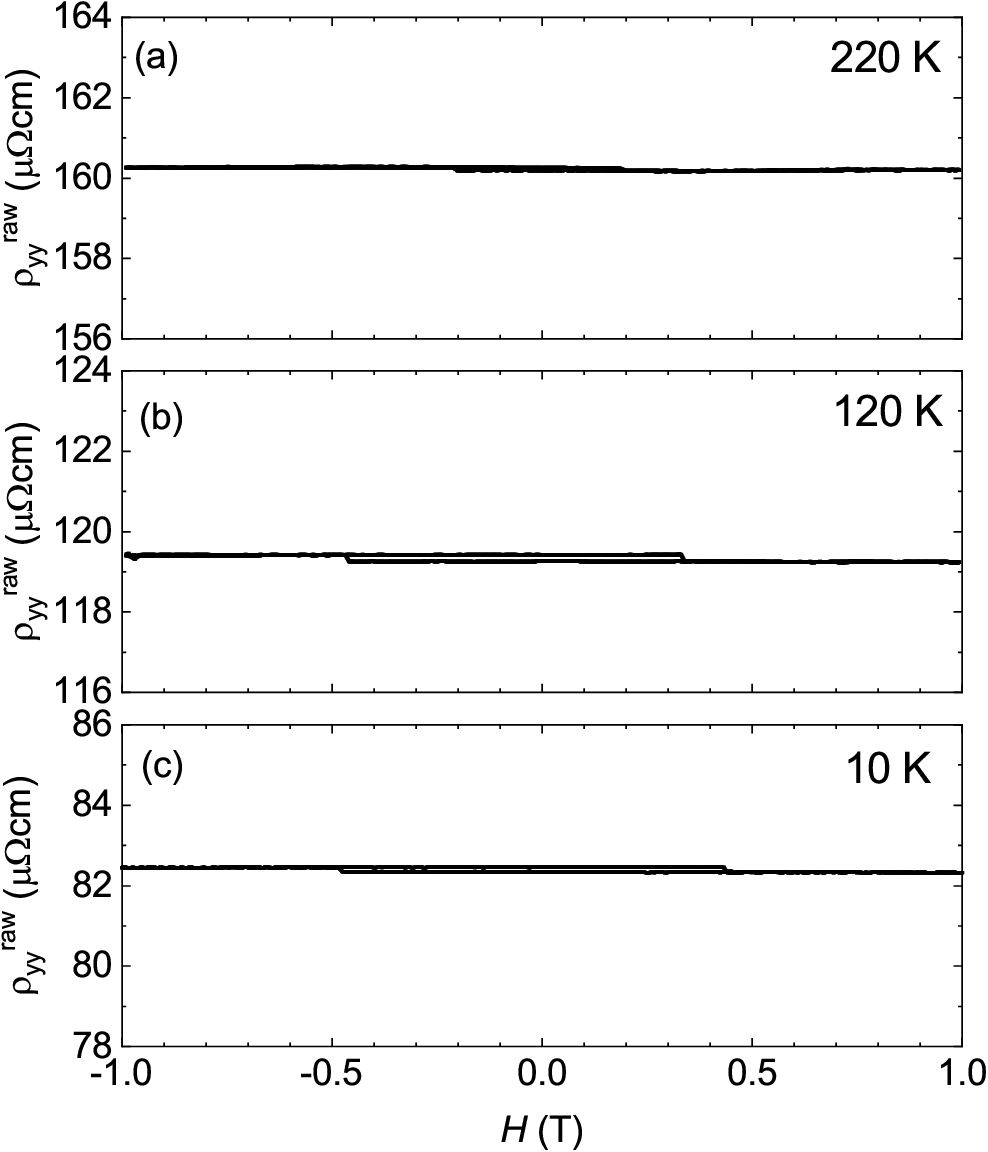}
\caption{The field dependence of the raw data of the electrical resistivity $\rho_{yy}$ for $H\parallel a$. The intrinsic $\rho_{yy}$ is insensitive to the magnetic field, although very small contribution from $\rho_{yz}$ was seen.}
\label{structure}
\end{figure}
%=========================================================

\clearpage 

\subsection{Crystal symmetry in the noncollinear AF state}

The symmetrical analysis clarified that the symmetry of NbMnP below $T_{\rm N}$ is $Pmn2_1$.
We investigated the crystal symmetry below $T_{\rm N}$ using single-crystal X-ray diffraction measurements, because it may accompany a structural deformation.
The results at 293 K, 200 K \cite{Matsuda1} and 120 K are shown in Table I\hspace{-1.2pt}I\hspace{-1.2pt}I.
Within the experimental resolution, there was no clear indication of lowering of the crystal symmetry from $Pnma$ to $Pmn2_1$.

\begin{table}[htb]
\caption{Structural parameters of NbMnP determined by single-crystal X-ray diffraction measurements at $T=293$ K, 200 K, and 120 K. Wyckoff positions of all atoms are $4c$.}
\label{t1}
\raggedright
\begin{center}
\begin{tabular}{ccccccc}
\hline
$T=293$ K& & & & & \\
\hline
Atom & $x$ & $y$ & $z$ & $Occ.$ &$U$(\AA$^2$)\\
\hline
Nb & 0.03102(5) & 0.25000 & 0.67215(5) & 0.968 & 0.00582(15)\\
Mn & 0.14147(9) & 0.25000 & 0.05925(8) & 1 & 0.0063(2)\\
P & 0.26798(15) & 0.25000 & 0.36994(13) & 1 & 0.0061(2)\\
\hline
\end{tabular}\\
orthorhombic ($Pnma$): $a$=6.1823(2) \AA, $b$=3.5573(2) \AA, $c$=7.2187(3) \AA, $R$=1.90\% \\
\vspace{3ex}
\begin{tabular}{ccccccc}
\hline
$T=200$ K& & & & & \\
\hline
Atom & $x$ & $y$ & $z$ & $Occ.$ &$U$(\AA$^2$)\\
\hline
Nb & 0.03123(6) & 0.25000 & 0.67187(5) & 0.969 & 0.00486(18)\\
Mn & 0.14152(9) & 0.25000 & 0.05945(8) & 1 & 0.0053(2)\\
P & 0.26790(15) & 0.25000 & 0.36949(14) & 1 & 0.0052(3)\\
\hline
\end{tabular}\\
orthorhombic ($Pnma$): $a$=6.1841(3) \AA, $b$=3.5504(2) \AA, $c$=7.2295(4) \AA, $R$=2.25\% \\
\vspace{3ex}
\begin{tabular}{ccccccc}
\hline
$T=120$ K & & & & & \\
\hline
Atom & $x$ & $y$ & $z$ & $Occ.$ &$U$(\AA$^2$)\\
\hline
Nb & 0.03152(5) & 0.25000 & 0.67160(4) & 0.971 & 0.0035(16)\\
Mn & 0.14134(7) & 0.25000 & 0.05962(6) & 1 & 0.0038(2)\\
P & 0.26761(12) & 0.25000 & 0.36932(11) & 1 & 0.0040(2)\\
\hline
\end{tabular}\\
orthorhombic ($Pnma$): $a$=6.1845(3) \AA, $b$=3.5446(2) \AA, $c$=7.2376(3) \AA, $R$=2.06\% \\
\end{center}
\end{table}

\newpage

\subsection{Consideration of the magnetization for four AF domains}

Occurrence of a net magnetization is understood naively from the breaking of the AF couplings of the Mn1--Mn4 and Mn2--Mn3 bonds.
As shown for $B_{2u}+B_{3g}$ (SA) in Table.~I, equivalency of the magnetic moments at four Mn sites is broken because of the symmetry-allowed components of the magnetic moments.
The magnetic moments for the respective domains are shown in Table~I\hspace{-1.2pt}V.
These configurations produce net magnetizations along the $a$-axis for all the domains even at zero field.
Directions of the net magnetization of the domain I and I\hspace{-1.2pt}I are the same, and they are the opposite from the domains I\hspace{-1.2pt}I\hspace{-1.2pt}I and I\hspace{-1.2pt}V.
This symmetry is maintained under $H \parallel a$, suggesting that two domains are selected.

When we applied magnetic fields along $H \parallel c$, further symmetry lowering occurs, resulting in four inequivalent Mn sites.
In this case, the averaged magnetizations along the $c$-axis differ between the domains (I and I\hspace{-1.2pt}I\hspace{-1.2pt}I) and (I\hspace{-1.2pt}I and I\hspace{-1.2pt}V).

\begin{table*}[htb]
\caption{The components of the symmetry-allowed magnetic moments for four domains of NbMnP. Induced moments by magnetic fields are not considered.}
\begin{center}
\begin{tabular}{ccccccc}\hline
$H =0$ and $H \parallel a$, $Pmn2_1$ & & & & & & \\
domain & &Mn1 & Mn2 & Mn3 & Mn4 & sum \\ \hline
I & & ($u$, 0, $v$) & ($u$, 0, $-v$) & ($-u'$, 0, $v'$) & ($-u'$, 0, $-v'$) & ($u_{\rm sum}$, 0, 0) \\
I\hspace{-1.2pt}I & & ($u$, 0, $-v$) & ($u$, 0, $v$) & ($-u'$, 0, $-v'$) & ($-u'$, 0, $v'$) & ($u_{\rm sum}$, 0, 0) \\
I\hspace{-1.2pt}I\hspace{-1.2pt}I & & ($-u$, 0, $v$) & ($-u$, 0, $-v$) & ($u'$, 0, $v'$) & ($u'$, 0, $-v'$) & ($-u_{\rm sum}$, 0, 0) \\
I\hspace{-1.2pt}V & & ($-u$, 0, $-v$) & ($-u$, 0, $v$) & ($u'$, 0, $-v'$) & ($u'$, 0, $v'$) & ($-u_{\rm sum}$, 0, 0) \\ \hline 
\end{tabular} \\
$u_{\rm sum}=2(u-u')$ \\ 
\vspace{3ex}
\begin{tabular}{ccccccc}\hline
$H \parallel c$ & & & & & & \\
domain & &Mn1 & Mn2 & Mn3 & Mn4 & sum \\ \hline
I & & ($u$, 0, $v$) & ($u'$, 0, $-v'$) & ($-u''$, 0, $v''$) & ($-u'''$, 0, $-v'''$) & ($u_{\rm sum}$, 0, $v_{\rm sum}$) \\
I\hspace{-1.2pt}I & & ($u$, 0, $-v$) & ($u'$, 0, $v'$) & ($-u''$, 0, $-v''$) & ($-u'''$, 0, $v'''$) & ($u_{\rm sum}$, 0, $-v_{\rm sum}$) \\
I\hspace{-1.2pt}I\hspace{-1.2pt}I & & ($-u$, 0, $v$) & ($-u'$, 0, $-v'$) & ($u''$, 0, $v''$) & ($u'''$, 0, $-v'''$) & ($-u_{\rm sum}$, 0, $v_{\rm sum}$) \\
I\hspace{-1.2pt}V & & ($-u$, 0, $-v$) & ($-u'$, 0, $v'$) & ($u''$, 0, $-v''$) & ($u'''$, 0, $v'''$) & ($-u_{\rm sum}$, 0, $-v_{\rm sum}$) \\ \hline 
\end{tabular} \\
$u_{\rm sum}=u+u'-u''-u'''$, $v_{\rm sum}=v-v'+v''-v'''$ \\
\end{center}
\end{table*}

\newpage

%\section{Asymmetric hysteresis for a different crystal}

%Hysteresis loops after ZFC and FC were measured using a different crystal.
%In this crystal, a sign change of $\rho_{yz}$ occurred successively through several steps.
%They are classified by three regions denoted as A, B, and C from the size of the change in $\rho_{yz}$.

%========FIGURE INSERTION=================================
%\begin{figure}[htb]
%\includegraphics[width=6.5cm]{sfig4.eps}
%\caption{The field dependence of the raw data of the electrical resistivity $\rho_{yy}$ for $H\parallel a$. The intrinsic $\rho_{yy}$ is insensitive to the magnetic field, although the very small contribution from $\rho_{yz}$ is seen.}
%\label{structure}
%\end{figure}
%=========================================================

\end{document}